\documentclass[12pt]{article}
\usepackage{amssymb}
\newtheorem{theorem}{Theorem}[section]
\newtheorem{proposition}{Proposition}[section]

\newtheorem{pelda}[theorem]{Example}

\def\tr{{\rm Tr}\,}

\def\L{\Lambda}
\def\vfi{\varphi}
\def\eps{\varepsilon}
\def\id{\mbox{id}}
\def\iK{{\cal K}}
\def\iH{{\cal H}}
\def\iA{{\cal A}}
\def\iB{{\cal B}}
\def\bbbn{\mathbb{N}}
\def\bbbc{\mathbb{C}}
\def\bbbz{\mathbb{Z}}
\def\<{\langle}
\def\>{\rangle}
\def\eps{\varepsilon}

\def\iAn{{\iA}^{\otimes  n}}
\def\iAv{{\iA}^{\otimes  \infty}}
\def\vphi{\varphi}

\def\prn{ \mathbb{P}(\iAn)}
\def\supp{\mbox{supp}}

\begin{document}
\ \vskip 1cm
\centerline{\LARGE Stationary quantum source coding}
\bigskip
\bigskip
\centerline{\Large D\'enes Petz\footnote{Supported by the Hungarian National 
Foundation for Scientific Research grant no. OTKA T 032662, e-mail: 
petz@math.bme.hu.} and
Mil\'an Mosonyi\footnote{E-mail: mosonyi@chardonnay.math.bme.hu.}}
\bigskip
\centerline{Mathematical Institute}
\centerline{Budapest University of Technology and Economics}
\centerline{H-1521 Budapest XI. Sztoczek u.\ 2, Hungary}
\bigskip
\begin{abstract}
In this paper the quantum version of the source coding theorem is obtained 
for a completely ergodic source. This result extends Schumacher's
quantum noiseless coding theorem for memoryless 
sources. The control of the memory effects requires some earlier results of 
Hiai and Petz on high probability subspaces. Our result is equivalently
considered as a compression theorem for noiseless stationary channels.
\end{abstract}

\section{Introduction}

Although it is difficult to define a discipline, to give some idea we
can say that the objective of quantum information theory is the
transmission and manipulation of information stored in systems obeying
quantum mechanics. A quantum channel has a source that emits systems in 
quantum states to the channel. For example, the source could be a laser
that emits individual monochromatic photons and the channel could be an 
optical fiber. The noisy signal output of the channel arrives at the 
receiver. In principle, there are two very different problems about 
quantum channels. The sender has a quantum system in an unknown state 
and wants to have the receiver to end up with a similar system in the
same state. In this case we speak of a pure quantum channel which has a 
quantum mechanical input and output. On the other hand, one might want 
to use quantum states to carry classical information, roughly 
speaking a sequence of zeros and ones. Now both the input and the output are
classical, however there is a quantum mechanical section inbetween. The classical
information is encoded into a quantum state and this is sent down the channel. 
The higher the channel noise is, the more redundant the encoding must be in order 
to restore the original signal at the reciever, where the quantum signal is 
converted into classical information. In this paper we do not deal with the problem 
how such a scheme can be realistically implemented; practical quantum encoding and
decoding requires sophisticated ability to manipulating quantum states. However,
we are interested in the amount of classical information getting through the channel
which is assumed to be noiseless. It was emphasized already by Shannon that
a computer memory is a communication channel. (Quantum or classical depends on
the type of the computer.) In an optimal situation the computer memory is free of 
any noise and this is the case we are concentrating on in the present paper. We 
want to consider rather general noiseless quantum channels (with possibly memory
effects but strong ergodic properties) and our aim is to discuss the quantum source
coding theorem. As a general reference on quantum information theory we
suggest the recent book \cite{N-Ch} but the really necessary definitions
are given below.
 
To each classical input message $x_i$ there corresponds a signal state $\vfi_i$ of 
the quantum communication system. The quantum states  $\vfi_i$ are functioning as
codewords of the messages. The signal states $\vfi_i$ could be pure and orthogonal
in the sense of quantum mechanics but for example in quantum cryptography 
nonorthogonal states are used intentionally in order to avoid eavesdropping. At
the moment we do not impose any condition on the signal states, they could be
arbitrary pure or mixed states. In the stochastic model of communication, one
assumes that each input message $x_i$ appears with certain probability. Let $p_{ji}$
be the probability that the message $x_i$ is sent and $y_j$ is recieved. The joint
distribution $p_{ji}$ yields marginal probability distributions $p_i$ and $q_j$
on the set of input and output messages. According to Shannon the mutual information
$$
I=\sum_{i,j} p_{ji}\log {p_{ji} \over p_i q_j }
$$
measures the amount of information going through the channel from Alice to Bob. Of
course, the relation of $I$ to the quantum encoding and decoding should be made
clear. This comes next.

The message $x_i$ has {\it a priori} probability $p_i$ and the mixed quantum state
of the channel is
$$
\vfi=\sum_i p_i \vfi_i .
$$
This might be considered as the statistical operator of the {\it mesagge ensemble},
for example when $\vfi_i$ is a pure state $|i\>\<i|$, then $\vfi=\sum_i p_i 
|i\>\<i|$ acts on the input Hilbert space $\iH$. The distribution of the output is 
determined by a measurement, which is nothing else but a physical word for 
decoding. To each output message there corresponds an obsevable $A_j$ on the
output Hilbert space $\iK$. It is customary to assume that $0 \le A_j$,
$\sum_j A_j=\id$ ($\id$ stands for the identity operator) and $p_{ji}=p_i \vfi_i(A_j)$.
The so-called {\it Kholevo bound} (\cite{Ho1}) provides an upper bound 
on the amount of information accessible to Bob in terms of von Neumann entropies:
$$
I \le S(\vfi)-\sum_i p_i S(\vfi_i) 
$$
(When $\lambda_1,\lambda_2, \dots$ are the eigenvalues of the statistical operator
of a quantum state $\psi$, then $S(\psi)=-\sum_k \lambda_k \log(\lambda_k)$.) In 
particular, if all signal states $\vfi_i$ are pure, then $S(\vfi_i)=0$ and
we have $I\le S(\vfi)$. In this way the von Neumann entropy gets an information
theoretical interpretation. Kholevo's bound is actually not very strong, it is 
attained only in trivial situations (\cite{OPW}).

The basic problem of communication theory is to maximize the amount of information
received by Bob from Alice. However, up to now this problem is not well-posed
in our discussion yet. Let us deal with messages of length $n$, they are 
$n$-term-sequences of $0$ and $1$. (So the size of this message set is $2^n$.)
For each message length $n$ we carry out the above procedure of coding and decoding
and the amount of information going through the channel is $I_n$. Since $I_n$
is presumably proportional to $n$, the good information quantity is $I_n/n$,
that is, the transmitted information per letter. Since Shannon's theory is not only
stochastic but asymptotic as well, we are going to let $n$ to $\infty$. In this
way we need to repeat the above information transmission scheme for each $n$. The
mesagge set, the input Hilbert space $\iH^{(n)}$, our coding, the channel state
$\vfi^{(n)}$, the output Hilbert space $\iK^{(n)}$ and the observables applied in 
the measurement are all depending on the parameter $n$.  

The subject of the present paper is faithful signal transmission, which bears the
name noiseless channel. In place of faithful transmission, one can think of 
information storage. In this case the aim is to use the least possible number of 
Hilbert space dimension per signal for coding. The new feature of the noiseless
channel we are studying is the memory effect. Mathematicaly this means that the
channel state (of the $n$-fold channel) is not of product type but we assume 
stationarity and good ergodic properties. In Section 2 we use the standard 
formalism of statistical mechanics to describe such a channel. It turns out that
the mean von Neumann entropy, familiar also from statistical mechanics, gives
the optimal coding rate. The proof of our main result, Theorems \ref{thmpoz} and
\ref{thmneg}, is similar to the proof
presented in \cite{JSch} for Schumacher's coding theorem, however instead of
typical sequences we use the high probability subspace of strongly ergodic
stationary states, a subject studied by Hiai and Petz in \cite{HP1}. We note
for the interested reader that most of the concepts used in the present paper
are treated in details in the monograph \cite{OP}.
 
\section{An infinite system setting of the source}

If $\iH$ is a finite dimensional Hilbert space then
$(A,B) \mapsto \tr(A^* B)$ defines an inner product on $\iB(\iH)$, so for
every linear functional $\vfi$ on $\iB(\iH)$ there exists a unique $D_{\vfi}
\in \iB(\iH)$ with the property $\vfi(A)=\tr(D_{\vfi} A)$. When
$\vfi$ is a state then $D_{\vfi}$ is the corresponding density matrix.
Let $X^n$ denote the set of all messages of length $n$. If $x^n \in X^n$ is a message
then a quantum state $\vfi(x^n)$ of the $n$-fold quantum system is corresponded with
it. The Hilbert space of the $n$-fold system is the $n$-fold tensor product 
$\iH^{\otimes n}$ and  $\vfi(x^n)$ has a statistical operator $D(x^n)$. If messages
of length $n$ are to be transmitted then our quantum source should be put in
the state $\vfi_n=\sum_{x^n} p(x^n) \vfi(x^n)$ with statistical operator
$D_n=\sum_{x^n} p(x^n) D(x^n)$, where $p(x^n)$ is the probability of the message
$x^n$. Since we want to let $n \to \infty$, it is reasonable to view all the 
$n$-fold systems as subsystems of an infinite one. In this way we can conveniently
use a formalism standard in statistical physics, see Chap. 15 of \cite{OP}. 

Let an infinitely extended system be considered over the lattice $\bbbz$ of integers.
The observables confined to a lattice site $k\in\bbbz$ form the selfadjoint part of 
a finite dimensional matrix algebra $\iA_k$, that is the set of all operators
acting on the finite dimensional space $\iH$. It is assumed that the
local observables in any finite subset $\L\subset\bbbz$ are those of the finite 
quantum system
$$
\iA_\L=\mathop{\otimes}_{k\in\L}\iA_k.
$$
The quasilocal algebra $\iA$ is the norm completion of the normed algebra
$\iA_\infty=\cup_{\L}\iA_\L$, the union of all local algebras $\iA_\L$ associated 
with finite intervals $\L\subset\bbbz$.

A state $\vfi$ of the infinite system is a positive normalized functional $\iA \to 
\bbbc$. It does not make sense to associate a statistical operator to a state
of the infinite system in general. However, $\vfi$ restricted to a finite dimensional
local algebra $\iA_\L$ admits a density matrix $D_\L$. We regard the algebra 
$\iA_{[1,n]}$ as the set of all operators acting on the $n$-fold tensor product space
$\iH^{\otimes n}$. Moreover, we assume that the density $D_n$ from the first part 
of this section is identical with $D_{[1,n]}$. Under this assumptions we call the
state $\vfi$ the state of the (infinite) channel. Roughly speaking, all the
states used in the transmission of messages of length $n$ are marginals of this 
$\vfi$. Coding, transmission and decoding could be well formulated using the states
$\vfi_n\equiv \vfi_{[1,n]}$. However, it is more convenient to formulate our setting
in the form of an infinite system, particularly because we do not want to assume
that the channel state $\vfi$ is a product type. This corresponds to the possibility
that our quantum source has a memory effect. 

The right shift on the set $\bbbz$ induces a transformation $\gamma$ on $\iA$. A 
state $\vfi$ is called {\it stationary} if $\vfi \circ \gamma=\vfi$. The state $\vfi$
is called {\it ergodic} if it is an extremal point in the set of stationary states.
Moreover, $\vfi$ is {\it completely ergodic} when it is an extreme point for every
$m \in \bbbn$ in the convex set of all states $\psi$ such that $\psi\circ \gamma^m 
=\psi$. By a {\it completely ergodic stationary quantum source} we simply mean a 
completely ergodic stationary state $\vfi$ of the infinite system $\iA$. Of course,
a stationary product state, corresponding to a memoryless channel, is completely
ergodic. The emphasis is put to other states here.

Below we show  an example of a completely ergodic stationary 
quantum source from the context of algebraic states. For the
details see the original paper \cite{HP2}.
 
\begin{pelda}
Let $\iA := M_3(\mathbb{C})$, $\iB :=M_2(\mathbb{C})$, moreover let
$\{E_{ij} \}_{i,j=1}^3$ be the usual matrix units of $M_3(\mathbb{C})$.
Set
$$
V_1:= \left[\matrix{ {1 \over \sqrt{2}} & 0 \cr 0  & 0 }\right],\quad
V_2:= \left[\matrix{ 0 & 0 \cr {1 \over \sqrt{2}} & 0 }\right],\quad
V_3:= \left[\matrix{ 0 & 1 \cr 0  & 0 }\right].
$$
Then $\sum_{i=1}^3 V_i^*V_i=I_{\iB}$. 

Let $\rho$ be a state on $\cal B$ with density matrix 
$$
\left[\matrix{ {2 \over 3} & 0 \cr 0  & {1 \over 3} }\right].
$$
Define $\Sigma:\iA \otimes \iB \to \iB$ by $\Sigma(E_{ij}\otimes x):= 
V_i^* x V_j$. It is easy to check that $\Sigma$ is a completely positive 
unital map and $\rho (\Sigma(I_{\iA}\otimes x))=
\rho(x),x\in \iB$. 

Then the algebraic state $\vfi$ generated by $(\iB,\Sigma, \rho)$ is given by 
$$
\vfi(E_{i_1 j_1}\otimes \dots \otimes E_{i_n j_n})=\rho(V_{i_1}^*\dots V_{i_n}^*V_{j_n}
\dots V_{j_1}).
$$
It is shown in \cite{HP2}  that $\vfi$ is completely ergodic. Of course, it is 
not a product state.
\end{pelda}

It is well-known in quantum statistical mechanics that due to the subadditivity
of the von Neumann entropy (proven first in \cite{LR} by Lieb and Ruskai) the limit
$$
\lim_{n\to +\infty} \frac{1}{n} S(\vfi_n)= \inf \frac{1}{n} S(\vfi_n)=:h
$$
exists for any stationary state and this quantity is called the {\it mean entropy} 
of $\vfi$. (See \cite{OP} for a textbook treatment of the subject or \cite{Pe} 
for some related properties of the mean entropy.)

\section{Source coding}

For a while we fix a message length $n$ and we denote by $d$ the dimension of the
Hilbert space $\iH$. Assume that our $n$-fold composite quantum system is operating
as a quantum source and emits the quantum states $D^{(1)}, D^{(2)}, \dots,
D^{(m)}$ with a-priory probabilities $p_1,p_2,\dots, p_m$. (Therefore the state
of the system is $D_n=\sum_i p_i D^{(i)}$.) By source coding we mean an association
$$
D^{(i)}\mapsto \tilde{D}^{(i)},
$$
where $\tilde{D}^{(i)}$ is some other statistical operator on the Hilbert space 
$\iH^{\otimes n}$. (This definition allows $D^{(i)}=D^{(j)}$ but
$\tilde{D}^{(i)}\ne \tilde{D}^{(j)}$, however in the coding constructed
in the proof of Theorem 1 this cannot happen.)

We denote by $\iK_n$ the subspace spanned by the eigenvectors 
corresponding to all nonzero eigenvalues of all statistical operators 
$\tilde{D}^{(i)}$, $1 \le i \le m$. The goal of source coding is to keep the 
dimension of $\iK_n$ to be small and to fulfil some fidelity criterium. 
(A mathematically demanding survey about quantum coding is the paper \cite{Ho}.)
The {\it source coding rate}
$$
\limsup_{n\to \infty} \frac{\log \dim (\iK_n)}{n} 
$$
expresses the resolution of the encoder in qubits per input symbol. (It is actually
more precise to speak about ``qunats'' per input symbol, but the difference is
only a constant factor.) 

The distortion measure is a number which allows us to compare the goodness or 
badness of communication sytems. The {\it fidelity} of the coding scheme was
introduced by Schumacher (\cite{Sch}):
$$
F:=\sum_{i} p_i \tr D^{(i)} \tilde{D}^{(i)},
$$
where $p_i$ is a probaility distribution on the input and $\tilde{D}^{(i)}$ is the 
density used to encode the density $D^{(i)}$. Note that $0\le F\le 1$ and $F=1$ if 
and only if $D^{(i)}=\tilde{D}^{(i)}$ are pure states.

First we present our positive source coding theorem for a completely ergodic
source. The result says that the source coding rate may approach the mean
entropy while we can keep the fidelity arbitrarily good.

\begin{theorem}\label{thmpoz}
Let $\iH$ be a finite dimensional Hilbert space, and
$\vfi$ be a completely ergodic state on $B(\iH)^{\otimes \infty}$.
Then for every $\eps, \delta>0$ there exists $n_{\varepsilon,\delta}
\in \bbbn$ such that for $n \ge n_{\varepsilon,\delta}$ there is
a subspace $\iK_n(\varepsilon,\delta)$ of $\iH^{\otimes n}$ such that
\begin{itemize}
\item[(i)] $\log \dim \iK_n(\varepsilon,\delta) < n(h+\delta)$ and
\item[(ii)] for every extremal decomposition $D_n=\sum_{i=1}^{m} p_i {D}^{(i)}$ one can 
find an encoding ${D}^{(i)}\mapsto \tilde{D}^{(i)}$ with density matrices 
$\tilde{D}^{(i)}$ supported in $\iK_n(\varepsilon,\delta)$ 
such that the fidelity $F:=\sum_{i=1}^{m} p_i \tr D^{(i)} \tilde{D}^{(i)}$ 
exceeds $1-\varepsilon$.
\end{itemize}
\end{theorem}

The negative part of the coding theorem tells that the source coding rate
cannot exceed the mean entropy when the fidelity is good.

\begin{theorem}\label{thmneg}
Let $\iH$ be a finite dimensional Hilbert space, and
$\vfi$ be a completely ergodic state on $B(\iH)^{\otimes \infty}$.
Then for every $\delta>0$ there exist $0 < \eta< 1$ and $n_{\delta}
\in \bbbn$ such that for $n \ge n_{\delta}$ 
\begin{itemize}
\item[(i)] for all subspaces $\iK_n$ of $\iH^{\otimes n}$
with the property
$\log$ dim $\iK_n \le n(h-\delta)$ and
\item[(ii)] for every decomposition $D_n=\sum_{i=1}^{m} 
p_i {D}^{(i)}$ and for
every encoding ${D}^{(i)}\mapsto \tilde{D}^{(i)}$ with density matrices 
$\tilde{D}^{(i)}$ supported in $\iK_n$, the fidelity
$F:=\sum_{i=1}^{m} p_i \tr D^{(i)} \tilde{D}^{(i)}$ is smaller than 
$\eta$.
\end{itemize}
\end{theorem}

The detailed proofs are given in the next section of the paper. Now we make
some comments on the fidelity $F$. It is possible that $F<1$ although
$D^{(i)}=\tilde{D}^{(i)}$. This fact might suggest to use another concept
of fidelity. Since $D^{1/2}\ge {D}$ holds for a density matrix,  we have
\begin{eqnarray*}
\tr {D_1}^{1/2}{D_2}^{1/2}&= & \tr {D_1}^{1/4}{D_2}^{1/2} {D_1}^{1/4}
\ge \tr {D_1}^{1/4}{D_2} {D_1}^{1/4}=\tr {D_2}^{1/2}D_1^{1/2} {D_2}^{1/2}
\\ & \ge &\tr {D_2}^{1/2}{D_1} {D_2}^{1/2}= \tr {D_1}{D_2}.
\end{eqnarray*}
This implies that
$$
F':=\sum_{i} p_i \tr \big[ D^{(i)}\big]^{1/2} \big[\tilde{D}^{(i)}\big]^{1/2}
\ge F.
$$
Both our positive and negative source
coding theorems hold if $F$ is replaced by $F'$. (In case of Theorem \ref{thmpoz}
this follows from the inequality $F' \ge F$ and in the proof of Theorem \ref{thmneg}
we will show $F' \le \eta $.)

\section{High probability subspace}

The proof of Shannon's original source coding theorem is based on the 
typical sequences (\cite{CsK}, Chap. 1). The quantum extension of this result 
obtained
by Schumacher still benefits from the classical result. When the channel state
is a product, the densities $D_n$ commute and simultanous diagonalization is 
possible. If the memory effects are present, then these densities do not
commute and in some sense we are in a really quantum mechanical non-commutative
situation. Nevertheless, the high probability subspace can be used but new 
techniques are required.

Let $\iK$ be a Hilbert space and $D$ be a density matrix on $\iK$. $D$ has a Schatten
decomposition $D=\sum_i \lambda_i |f_i\>\<f_i|$, where $|f_i\>$'s are  eigenvectors
and the eigenvalues $\lambda_i$ are numbered decreasingly: $\lambda_1\ge \lambda_2 
\ge \dots $. Choose and fix $0 < \eps <1$. Let $n(\eps)$ be the smallest integer
such that
$$
\sum_{i=1}^{n(\eps)} \lambda_i \ge 1 -\eps\, .
$$ 
The subspace $HP(D,\eps)$ spanned by the eigenvectors $|f_1\>,\dots, |f_{n(\eps)}\>$
is called the {\it high probability subspace} corresponding to the level $\eps$. 
Note that $HP(D,\eps)$ is not completely well-defined, if there are 
multiplicities in the spectrum of $D$, then the Schatten decomposition is 
not unique. However, the dimension $n(\eps)$ of $HP(D,\eps)$ is determined. 
The term ``high probability subspace'' is borrowed from the monograph \cite{Gr} 
and its role in macroscopic uniformity was discussed in \cite {HP3}.

In the following, $\vfi$ will be a completely ergodic state on $\iAv$.
For $\eps \in$ (0,1) let 
$$
\beta_{\eps,n} : = \inf\{\log \tr_n (q)) \colon q \in \prn , \vfi_n(q) 
\geq  1-\eps\},
$$
where $\prn$ denotes the set of projections of $\iAn$. ($\exp \beta_{\eps,n}$ is
the dimension of the high probability subspace.) It was shown in 
\cite{HP1} (and formulated in terms of relative entropy) that
\begin{eqnarray}
\limsup_{n \to +\infty} \frac{1}{n}\beta_{\varepsilon,n} &\leq & h, \label{(1)}\\
\liminf_{n \to +\infty} \frac{1}{n}\beta_{\varepsilon,n} & \geq &
\frac{1}{1-\varepsilon} h -\frac{\eps}{1-\eps} \log d. \label{(2)}
\end{eqnarray}
From this one can deduce the following

\begin{proposition}\label{thmhp} For every positive $\delta$ 
\begin{itemize}
\item[(i)] and for every positive $\varepsilon$
there exists $N_{\varepsilon, \delta} \in \bbbn$ 
such that for every $n>N_{\varepsilon,\delta}$ there exists a projection 
$q_n(\eps,\delta)$ in $\iAn$ such that
$$
\log(\tr_n (q_n(\eps,\delta)))<n(h+\delta)\quad \hbox{and}\quad \vfi_n(q) \geq 1-\varepsilon,
$$
\item[(ii)] there exists $1 > \eta >0$ and $N_{\delta} \in \bbbn$ such
that for every $n>N_{\delta}$ and for every projection $q$ in $\iAn$
$$
\log(\tr_n (q))\le n(h-\delta),
$$
implies $\vfi_n(q)\le \eta$.
\end{itemize}
\end{proposition}

Part (i) of the Proposition is a plain reformulation of (\ref{(1)}). In order to see
(ii) we first note that 
$$
\frac{1}{\eta}h-\frac{1-\eta}{\eta}\log d \to h \quad \mbox{as}\quad
\eta\to 1\, .
$$
Hence given $\delta>0$ we choose $0<\eta<1$ such that
$$
\frac{1}{\eta}h-\frac{1-\eta}{\eta}\log d > h-\delta\,.
$$
Next we replace $1-\eta$ by $1-\delta$ in (\ref{(2)}):
\begin{equation}\label{eps}
\liminf_{n \to +\infty} {1 \over n}\inf\{\log \tr_n (q)) \colon 
q \in \prn , \vfi_n(q) 
\geq  1-\eta\} \ge {1 \over \eta}h-{1-\eta \over \eta}\log d >h-\delta.
\end{equation}
In this way we arrived at (ii).

Next we prove the source coding theorem.

{\it Proof of Theorem \ref{thmpoz}}:
Use part (i) of the Proposition and set $q_n:=q_n(\eps/2,\delta)$ ,
$\iK_n(\varepsilon,\delta):=\hbox{Ran}\, q_n$, where
$n>n(\eps,\delta):=N_{\eps/2,\delta}$. Given an extremal decomposition $D_n=
\sum_{i=1}^{k} p_i D^{(i)}$, that is $D^{(i)}=|x_i\>\<x_i|$ for some
vectors $x_i$, we construct the coding densities $\tilde{D}^{(i)}$. Let 
$$
\tilde{x_i}:=\frac{q_n x_i}{\|q_n x_i \|}, \quad \alpha_i:=\|q_n x_i \|,
\quad \beta_i:=\|(I-q_n) x_i \|
$$
and let $x$ be any unit vector such that $q_n x=x$. Then we set 
$$
\tilde{D}^{(i)}:=\alpha_i^2 |\tilde{x_i}\>\<\tilde{x_i}|+\beta_i^2|x\>\<x|.
$$
Since $\tilde{x_i},x \in \iK_n(\varepsilon, \delta)$, we have $\supp 
\tilde{D}^{(i)} \subset \iK_n(\varepsilon, \delta)$. Furthermore,
\begin{eqnarray*}
\tr {D}^{(i)}\tilde{D}^{(i)}&=&\<x_i,\tilde{D}^{(i)}  x_i\>=
\alpha_i^2 |\<x_i|\tilde{x_i}\>|^2+\beta_i^2 |\<x_i,x\>|^2 \\ &\geq&
\alpha_i^2 |\<x_i|\tilde{x_i}\>|^2 = \alpha_i^4 \geq 2\alpha_i^2-1\\ &=&
2\tr q_n {D}^{(i)}-1.
\end{eqnarray*}
We need to sum over $i$:
$$
\sum_i p_i \tr {D}^{(i)}\tilde{D}^{(i)}\ge \sum_i p_i \big(2\tr q_n 
{D}^{(i)}-1\big)=2\tr D_n q_n-1=2 \vfi_n (q_n)-1 \ge 1 - \eps.
$$

{\it Proof of Theorem \ref{thmneg}}:
For the given $\delta$ we choose $\eta$ and $n(\delta)$
according to the Proposition. Let $q$ be the projection onto the subspace 
$\iK_n$. We want to use the Schwarz inequality in the form
$$
\Big|\sum_i p_i \tr x_i y_i\Big| \le
\Big[\sum_i p_i \tr x_i^* x_i\Big]^{1/2} 
\Big[\sum_i p_i \tr y_i^* y_i\Big]^{1/2}
$$
for $x_i=[D^{(i)}]^{1/2} q$ and $y_i=[\tilde{D}^{(i)}]^{1/2}$. 
Since $[\tilde{D}^{(i)}]^{1/2}=q[\tilde{D}^{(i)}]^{1/2}$ follows from 
the hypothesis, we have 
\begin{eqnarray*}
F'&=&\sum_{i=1}^{m} p_i \tr [D^{(i)}]^{1/2}[ \tilde{D}^{(i)}]^{1/2}
= \sum_{i=1}^{m} p_i \tr [D^{(i)}]^{1/2}q[ \tilde{D}^{(i)}]^{1/2}\\
&\le & \Big[\sum_{i=1}^{m} p_i \tr D^{(i)} q \Big]^{1/2}
\Big[\sum_{i=1}^{m} p_i \tr \tilde{D}^{(i)} \Big]^{1/2}\\
&=& \vphi_n (q)^{1/2} \le \sqrt{\eta}.
\end{eqnarray*}
This estimate completes the proof.

It is known  that for strongly mixing algebraic states (\cite{HP2})
and for ergodic Gibbs states (\cite{HP3})
$$
\lim_{n \to +\infty} \frac{1}{n} \beta_{\eps,n}=h
$$ 
and in this case the negative part of the coding theorem holds in a stronger
form:
For every $\eps, \delta>0$ there exists $n_{\varepsilon,\delta}
\in \bbbn$ such that for $n \ge n_{\varepsilon,\delta}$ 
for all subspaces $\iK_n$ of $\iH^{\otimes n}$ 
with the property $\log$ dim $\iK_n < n(h-\delta)$ and
for every decomposition $D_n=\sum_{i=1}^{m} 
p_i {D}^{(i)}$ and for
every encoding ${D}^{(i)}\mapsto \tilde{D}^{(i)}$ with density matrices 
$\tilde{D}^{(i)}$ supported in $\iK_n(\varepsilon,\delta)$, the fidelity 
$F:=\sum_{i=1}^{m} p_i \tr D^{(i)} \tilde{D}^{(i)}$ is smaller than 
$\varepsilon$.

There is a seemingly slight difference between the two theorems. The 
statistical operator $D_n$ has an extremal decomposition in the first
one and arbitrary decomposition in the second. The difference between
the pure and mixed message ensemble is discussed in the recent paper 
\cite{jozsa}.

\section{Discussion}
In this paper a theory of quantum source coding subject to a fidelity criterion
or quantum data compression is presented. The minimum of the source coding
rate is studied under the conditions that Schumacher's fidelity must exceed
$1-\varepsilon$ and the quantum mechanical state of the channel has a strong
ergodic property. This latter condition allows many states with memory effect.
For the mathematical model and in the proof of the main result techniques
of quantum statistical mechanics are used. We prove that the minimal source
coding rate is the mean entropy of the channel state, and, to some extent,
it is independent of the message ensemble.  

\section{Acknowledgement}
The authors thank to Prof. O.E. Barndorff-Nielsen for an invitation to a
workshop held at MaPhySto, to Prof. A. Holevo and Dr. S. Furuichi for 
comments on the first draft of the paper.

\end{document}